\documentclass[aps,prc,twocolumn,superscriptaddress]{revtex4-2}

\usepackage{graphicx}
\usepackage{amsmath,amssymb}
\usepackage{hyperref}
\usepackage{bm}
\usepackage{siunitx}
\usepackage{bm}
\usepackage{braket}
\usepackage{physics}
\usepackage{xfrac}
\usepackage{blindtext}
\usepackage{tikz}
\usepackage{comment}

\begin{document}
	
	\title{ Resonant Nuclear Fusion at Second Order }
	
	\author{Harishyam Kumar\footnote{hari@iitk.ac.in}}
	\author{Pankaj Jain$^{2}$\footnote{pkjain@iitk.ac.in}}
	\affiliation{Physics Department, Indian Institute of Technology, Kanpur, 208016, India\\
		$^2$ Department of Space, Planetary and Astronomical Sciences \& Engineering, Indian Institute of Technology, Kanpur, 208016, India}
	
	
	\begin{abstract}
	We study the possibility of low energy nuclear fusion assisted by a low
	energy resonance. We use a simple potential model and replace  
	repulsive Coulomb barrier with a step potential barrier. This is convenient 
	since it allows us to perform analytic calculations and allows better control
	on approximations. The fusion process involves a proton and another nucleus.
	We first consider a process in which this transition takes place by emission
	of a single photon at first order in perturbation theory. At this order we find a
	very large cross section over a very narrow range of initial momenta, 
	corresponding to the resonance width. The cross section rapidly drops
	to very small values as we move away from resonance. In any
	experimental situation, the rate for this process would be negligible since
	it is practically impossible to have a significant number of initial
	state particles at precisely the resonant energy.  
	We next consider another process, which involves emission of two photons
	and gets dominant contribution at second order in perturbation theory. 
	In this case the initial state energy is taken to be larger than the resonance
	energy. We find a significant cross section, provided 
	the initial energy is
	larger than the resonant energy. 
	The exponential suppression factors cancel out in this calculation. 
	The experimental signature for this process
	is two photon emission within very narrow range of energies. One of the photon
	would be emitted at low energy of order of the initial state energy and
	the second would correspond to the nuclear energy.
	
	\end{abstract}
	
	\maketitle
	
	\section{Introduction}
There are many experimental claims that low energy nuclear reactions (LENR) may occur at observable rates even at relatively low energies of the order of few eV or even smaller \cite{doi:10.1002/9781118043493.ch41,doi:10.1002/9781118043493.ch42,doi:10.1002/9781118043493.ch43,biberian2020cold,PhysRevC.78.015803,StormsCS2015,McKubre16,Cellani19,Mizuno19,SRINIVASAN2020233}. Theoretically the nuclear fusion rates are expected to be exceedingly small at such energies due to the prohibitive Coulomb barrier. There exist many theoretical models, such as,
electron screening \cite{PhysRevC.101.044609,assenbaum1987effects,ichimaru1993nuclear}, correlated states \cite{articleVy, PhysRevAccelBeams.22.054503}, electroweak interactions \cite{Widom10}, formation of clusters of nuclear particles \cite{SPITALERI2016275}, relativistic electrons in deep orbits \cite{Meulenberg19} and phonon induced reactions \cite{Hagelstein15}, which attempt to explain this phenomenon. There is currently no agreement
among physicists about the validity of these models and a critical review is given in \cite{Chechin_1994}.

Another approach to the problem is through the second order perturbation theory \cite{merzbacher1998quantum,sakurai1967advanced}, and has been pursued in 
\cite{PhysRevC.99.054620, Jain2020, Jain2021, ramkumar2022, Kumar:2023}.  
In this case there are two interaction vertices. The low energy initial state undergoes a transition 
to the intermediate state at the first vertex by transfering energy and momentum to a particle in 
the medium \cite{PhysRevC.99.054620} or by emitting particles, such as photons \cite{Jain2020, Jain2021, ramkumar2022, Kumar:2023}. 
The intermediate state makes a transition to the final state at the second vertex. 
The amplitude corresponding to the first vertex is controlled by distance scales of order atomic units
while the second interaction gets dominant contributions from nuclear distance scale. It has been argued that due to the contribution from intermediate 
states of
arbitrarily high energy, the Coulomb
barrier may not be a major issue \cite{Jain2020, Jain2021}. However,  
despite the existence of such high energy states, it has been found that
rate for such a process in very small in most cases \cite{Jain2020, Jain2021, ramkumar2022, Kumar:2023}. One generally finds that contributions from different
high energy intermediate states tend to cancel one another. 
It has been suggested that this cancellation may be avoided in some special cases.
These invoke presence of a resonance in the nuclear spectrum 
\cite{ramkumar2024low,Jain:2024wbe} and specific forms of interaction \cite{Kumar:2023}. 

In this paper, we study the LENR process in the presence of a low energy
resonance. This is a state with energy eigenvalue $E>0$, whose eigenfunction 
takes very large values
inside the nuclear region and is very small outside. We are interested 
in such states with small energy of order of atomic energy, i.e. eV or tens of eV. 
For such energies, we expect all states to be
highly suppressed in the nuclear region. However, depending on the potential, 
the wave function may take very large values inside the nuclear region
for a very narrow range of energy values. 
We clarify that we are considering a state in the region where energy 
takes a continuous range of values.
We are interested in fusion of a proton
with a nucleus X to form a nucleus Y. 
For simplicity, we assume the mass of X is very large. In the case of 
finite mass, we only need to also include the center of mass motion
which does not make any essential change in the calculation.
Furthermore, we replace the repulsive Coulomb potential with a simple 
step potential. This allows us to perform analytic
calculations which permit control 
over approximations. As we shall argue, our
results are general and not limited by this simplifying assumption.

We first perform the calculation for a process in which the transition
involves only one photon emission. This is discussed in section \ref{sec:first}. The initial state in this case is 
taken to be the resonant state. This process gets dominant contribution 
at first order in perturbation theory. Although 
this process is very interesting, 
its practical applications are limited. This is because it is nearly 
impossible to prepare an initial state with precisely the energy corresponding
to resonance.  

We next consider a different process which involves emission of two photons.
As discussed in section \ref{sec:second}, this gets dominant contribution at second order in perturbation
theory. In this case, the initial state is taken to be any state with 
energy $E_i$ greater than the resonant energy $E_R$. At the first vertex 
the initial state makes a transition to another state with energy eigenvalue
$E_n$ by emission of a 
photon. The dominant contribution to this amplitude is obtained by $E_n$
close to the resonant energy. At the second vertex the intermediate
state makes a transtion to the final nuclear state. The process proceeds
by energy and momentum remaining conserved at each vertex. 
\section{Potential Model}
We consider a spherically symmetric potential model of the form,
\begin{equation}\label{eq:1}
	V(r) =
	\begin{cases}
		-V_0 & \text{$0\leq r <L_n$ }\\
		V_1 & \text{$L_n \leq r < L_b$}\\
		0 & \text{$ r > L_b$}
	\end{cases}    
\end{equation}
where $V_0$ corresponds to the nuclear potential and 
$V_1$ is our simplified model for the tunnelling barrier. 
The schematic diagram of potential is shown in Fig.1.
Here the deep well corresponds to the nuclear potential while the
step represents our model barrier potential. Our choice of potential 
parameters are given in section \ref{sec:results}. 
\begin{figure}[h]
	\centering
	\includegraphics[ clip,scale=0.70]{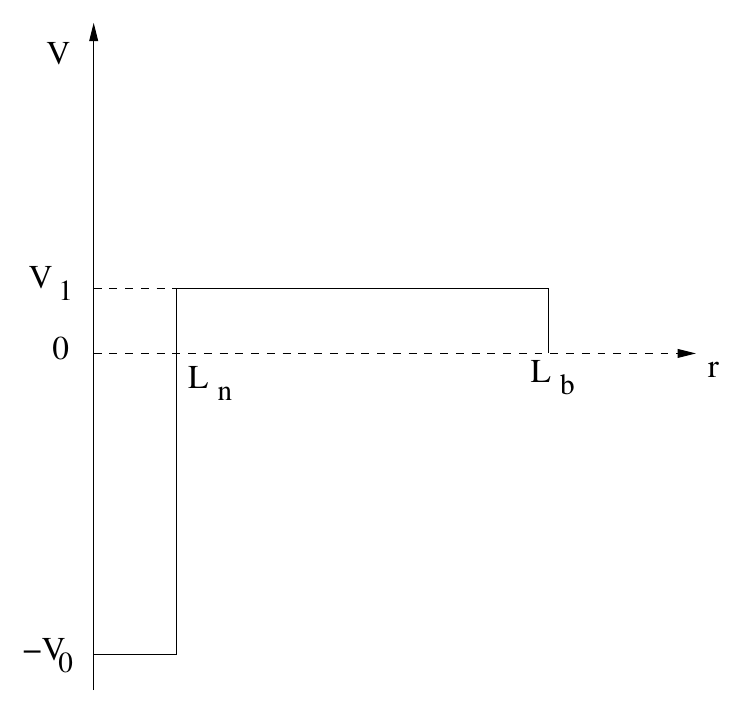}
	\caption{\label{fig:potential} Schematic illustration of model potential: Here
		the deep well corresponds to the nuclear potential with $L_n$ equal to the
		nuclear length scale. The potential step of height $V_1$ represents the
		repulsive potential with $L_b>> L_n$. 
	}
\end{figure}

We are interested in obtaining the wave functions of the initial
state p-X system. 
Here we focus on the relative coordinate $\vec{r}$ since  
the centre of the mass coordinate does not play any essential role
in the fusion analysis and, for simplicity, we take the X particle
mass to be very large.
The initial and final
states are taken to be $l=0$ and $l=1$ respectively. For the second
order process, the intermediate state also is taken to be $l=0$.   
The $l=0$ wave function can be expressed as, 
\begin{equation}
	\psi(r)= Y_0^0 \frac{U(r)}{r}
\end{equation}
where 
$Y_0^0=1/\sqrt{4 \pi}$.
The wave function $U(r)$ 
for energy eigenvalue $E>0$ 
can be expressed as,
\begin{equation}
	U(r) =
	\begin{cases}
		\frac{1}{N} \sin{k_1 r} & \text{$0\leq r < L_n$}\\
		\frac{1}{N} [ B e^{-\kappa_2 r}+C e^{\kappa_2 r}] & \text{$L_n\leq r <L_b $}\\
		\frac{1}{N} [ D \sin{k r }+F \cos{k r }] & \text{$r > L_b$}
	\end{cases}       
\end{equation}
where,
\begin{equation} 
	\begin{split}
		\hbar k_1 & =\sqrt{2 m_p(V_0+E)} \\
		\hbar \kappa_2 & =\sqrt{2 m_p(V_1-E)}\\
		\hbar k &=\sqrt{2 m_p E} 
	\end{split}
	\label{eq2:k_values}
\end{equation}
Here $N=\sqrt{D^2+F^2}\, \sqrt{L/2}$ is the normalization constant and
$L$ is the total length scale which is taken to infinity at the end of
the calculation.
The coefficients are given by, 

\begin{equation} \label{eq:coeff}
	\begin{aligned}
		B & =\frac{e^{\kappa_2 L_n} (\kappa_2 \sin (k_1 L_n)-k_1 \cos (k_1 L_n))}
		{2 \kappa_2} \\
		C & =\frac{e^{-\kappa_2 L_n} (\kappa_2 \sin (k_1 L_n)+k_1 \cos (k_1 L_n))}{2 \kappa_2}\\
		D &=	(B e^{-\kappa_2 L_b}+ C e^{\kappa_2 L_b}) \sin(k L_b) \\
		&\quad- \frac{\kappa_2}{k}(B e^{-\kappa_2 L_b}- C e^{\kappa_2 L_b}) \cos(k L_b)\\
		F &=	(B e^{-\kappa_2 L_b}+ C e^{\kappa_2 L_b}) \cos(k L_b) \\
		& \quad + \frac{\kappa_2}{k}(B e^{-\kappa_2 L_b}- C e^{\kappa_2 L_b}) \sin(k L_b)
	\end{aligned}
\end{equation}

In our analysis we are interested in the contribution from the resonance 
state. This state corresponds to energy $E>0$ at which the coefficient $C
\approx 0$. This corresponds to the condition,
\begin{equation}
	\tan(k_1L_n) \approx -{k_1\over \kappa_2}
\end{equation}
We denote the energy of the state at which this condition is exactly satisfied
by $E_R$.
In this case, the factor $D^2+F^2$ in the normalization constant,
given by

\begin{equation}
	\begin{aligned}
	D^2 + F^2 &= \left(Be^{-\kappa_2L_b} + C e^{\kappa_2L_b} \right )^2\\
	&\quad+ \left({\kappa_2\over k}\right)^2 
	\left(-Be^{-\kappa_2L_b} + C e^{\kappa_2L_b} \right )^2 \,,
\end{aligned}
\end{equation}
becomes very small and the wave function becomes 
very large within nuclear dimensions and
exceedingly small at larger distances. This is a meta-stable nuclear state
which is expected to have a very small width due to its very low energy.
The width is essentially controlled by the tunnelling factor 
$\exp(-\kappa_2L_b)$.
In our analysis we are interested in the inverse problem of a particle 
which is outside the nuclear well making a transition inside the nucleus.
Since the wave function is very small at large distances we expect that there would be very few particles within the small width corresponding to this state. We analyze this
in the next section by computing the cross section and reaction rate at 
first order. In the follow up section we compute the cross section for a 
second order process.

\section{Cross section at first order}
\label{sec:first}

In this section we consider a first order process in which a proton 
$p$ undergoes fusion with another nucleus $X$ 
with emission of a photon to form the final nucleus $Y$.
The process can be expressed as
\begin{equation}
	p + X \rightarrow Y + \gamma
	\label{eq:process1}
\end{equation}
The total Hamiltonian for the system can be written as,
\begin{equation}
	H=H_0+H_I
\end{equation} 
where $H_0$ is unperturbed Hamiltonian and $H_I$ perturbative  Hamiltonian.  We can write $H_0$ as,
\begin{equation}
	H_0=K_1+K_2+V(r)
\end{equation} 
where $K_1$ and $K_2$ are the kinetic energies of H and X nuclei respectively and $V(r)$ is the potential between them. 
As explained earlier, we take the mass of $X$ to be very large and ignore its
kinetic energy and focus on the relative motion. 
The interaction Hamiltonian
is given by,
\begin{equation}\label{eq:IH}
	H_I(t)=-\frac{Z_pe}{c m_p} \vec{A}(\vec{r}_1,t)\cdot \vec p_1-\frac{Z_Xe}{c m_X} \vec{A}(\vec{r}_2,t)\vec p_2+\frac{e\hbar g_p}{2 m_1 c}\vec{\sigma}.\vec{B}+...
\end{equation}
where $Z_p$ and $m_p$ are the atomic number and mass of the p
and $Z_X$ , $m_X$ are the corresponding quantities for X. The coordinates
of p and X are denoted by 
$\vec{r_1}$ and $\vec{r_2}$ respectively and their corresponding momenta 
are denoted by $\vec p_1$ and $\vec p_2$. 
Furthermore, $\vec{\sigma}_i$ are the Pauli matrices, $g_p$ the proton g factor, $\vec{A}$ is vector potential of photon field,
\begin{equation}
	\vec{A}(\vec{r},t)=\sum_{\vec{k}}\sum_{\beta}c\sqrt{\frac{\hbar}{2 \omega V}}\left[a_{\vec{k},\beta}(t)\vec{\epsilon}_\beta e^{i \vec{k}.\vec{r}}+ a^\dagger_{\vec{k},\beta}(t)  \vec{\epsilon}_\beta^* e^{-i \vec{k}.\vec{r}}\right]\,,
\end{equation} 
$\vec{B}=\vec{\nabla}\times \vec{A} $ is the magnetic field of the photon, 
$\epsilon_\beta$ is  polarization vector, $\vec k$ is the wave vector, 
$\omega$ is the frequency and $V$ the total volume, taken to be infinity at 
the end of the calculation.

We are interested in a transition from $l=0$ to $l=1$ state with emission of
a photon. There is no change in spin in this transition. The 
dominant contribution comes from 
the first term of Eq. \ref{eq:IH}. The second term also contributes but 
gives a relatively small contribution compared to the first term and 
does not lead to any essential change in our results. 
The photon wave number and frequency are denoted by $\vec k_\gamma$
and $\omega_\gamma$ respectively. 
We denote the polar coordinates of the unit vector $\hat k_{\gamma}$ by $\theta_\gamma$ and $\phi_\gamma$. The polarization vectors are denoted by $\vec \epsilon_1$ and $\vec \epsilon_2$. 
These three unit vectors are given by,
\begin{equation}
	\hat k_{\gamma} = \cos\theta_\gamma\hat z + \sin\theta_\gamma\left[\cos\phi_\gamma\hat x
	+\sin\phi_\gamma \hat y\right]
	\label{eq:kgamma}
\end{equation}
\begin{eqnarray}
	\vec\epsilon_1 &=& -\sin\theta_\gamma\hat z + 
	\cos\theta_\gamma\left[\cos\phi_\gamma\hat x +\sin\phi_\gamma \hat y\right] \nonumber\\
	\vec\epsilon_2 &=& -\sin\phi_\gamma\hat x +\cos\phi_\gamma \hat y 
	\label{eq:eps1_2}
\end{eqnarray}
We are only interested in order of magnitude estimate and hence confine ourselves to a particular polarization vector, which is taken to be
$\vec\epsilon_2$. The contribution from the remaining polarization vector
will add incoherently and hence produce a change of order unity which will
not change the result qualitatively.
The resulting matrix element is found to be,
\begin{align}
	\bra{f}H_I(t)\ket{i} &= -e^{i\omega_\gamma t}ie \sqrt{\frac{1}{2V \omega_\gamma\hbar}} (E_f-E_i) \bra{f} \vec\epsilon_2\cdot\vec r\ket{i}\nonumber\\
	& =	-e^{i\omega_\gamma t}ie \sqrt{\frac{1}{12 V \omega_\gamma\hbar}}\,
	(E_f-E_i)\nonumber\\
	&\quad \quad \times (\sin\phi_\gamma+i\cos\phi_\gamma) I_a 
\end{align}
where
\begin{equation}\label{eq:I2}
	I_a=\int dr'\ U_f^*\ r'\ U_i.
\end{equation}

The reaction rate can be expressed as,
\begin{equation}
	{dP\over dt} = {1\over \Delta T} \int dE_\gamma\rho_1
	|\langle f|T(t_0,t)|i\rangle|^2
\end{equation}
where $E_\gamma=\hbar \omega_\gamma$ is the photon energy and $\rho_1$ the photon
density of state factor. We obtain
\begin{equation}
	{dP\over dt} = {\alpha\omega_\gamma\over 3\hbar^2 c^2} \left(E_f-E_i
	\right)^2|I_a|^2
\end{equation}
The corresponding cross section is given by, $\sigma=(dP/dt)/(N_2F(E_i))$, where
$N_2=1$ is the number of target particles, $F(E)=v/V$ is the flux of incident particles and
$v= \sqrt{2E_i/m_p}$.
The important point is that the cross section is very large when $E_i
\approx E_R$. This is due to the factor $\exp(2\kappa_2L_b)$ contributed by
$1/N_i^2$, where $N_i$ is the normalization constant of the initial state.
However, we argue that despite the large cross section, this does not lead to
a large reaction rate in a physical experiment. This is due to the fact that
the cross section is large only over a very narrow range of energy values 
corresponding to the width of this state. 
The number of particles in this energy range are negligible in any experimental
situation and hence we will observe negligible reaction rate. 
The reaction rate may be enhanced by increasing the width of the resonance
by applying an external potential. This may be possible by some means, 
such as an applied electric or magnetic field. 
We do not pursue this in the
present paper and postpone this to future research.
Alternatively, we may 
fine tune the initial state to have precisely the resonance energy. However,
this appears to be nearly impossible.

\section{Cross section at second order}
\label{sec:second}

In this section we consider a process which gets it leading order contribution
at second order in perturbation theory. The reaction can be written as,
\begin{equation}
	p + X \rightarrow Y + \gamma(\omega_1) + \gamma(\omega_2)
	\label{eq:process2}
\end{equation}
We again assume presence of a low energy resonance. The target $X$ is taken
to be at rest and the incident particle $p$ has energy $E_i> E_R$. 
Due to two photon emission, the leading order contribution is obtained at
second order in perturbation theory, with one photon emitted at each 
vertex. The initial proton state is chosen such that, 
$l = 0$ and $S = 1/2$ with $S_z = -1/2$. We are interested in a transition to 
intermediate state with a spin flip,
with $S_z = -1/2$ going 
to $S_z = 1/2$ and no change in $l$.  
The dominant contribution comes from the magnetic interaction, the last term of interacting Hamiltonian eq. \ref{eq:IH}. 
The resulting matrix element can be written as,
\begin{equation}\label{eq:im}
		\begin{aligned}
	\bra{n}H_I(t)\ket{i}=	&\, \frac{i\ e\ \hbar\ g_p\ k_{\gamma 1}}{2m_p\sqrt V}\sqrt{\frac{\hbar}{2\omega_1}}\\[1mm]
	&\times\, \bra{n}\vec \sigma \cdot(\hat k_{\gamma 1}\times\vec \epsilon_1^*)\ a^\dagger\ e^{-i\ \vec k_{\gamma 1}\cdot\vec r+i\omega_1t}\ket{i}
\end{aligned}
\end{equation}
where $\vec k_{\gamma 1}$ is the photon wave vector and $\omega_1$ its 
frequency.
There is another contribution proportional to $1/m_X$, which is relatively small in the limit $m_X>>m_p$ and hence has been neglected \cite{Jain2020}. Furthermore, 
we have approximated the factor $m_X/(m_p+m_X)$ to be unity.
The photon wave and polarization vectors are given by equations similar to 
Eq. \ref{eq:kgamma} and \ref{eq:eps1_2}. We denote these vectors as
$\vec k_{\gamma 1}$, $\vec\epsilon_{\gamma 1}$ and $\vec\epsilon_{\gamma 2}$
respectively.
In our analysis we confine ourselves to final state photon polarization to be
$\vec\epsilon_1$. We point out that  
the contribution from $\vec\epsilon_2$ will add 
incoherently and can only change the result by a factor of order unity 
and here we are only interested in an order of magnitude estimate.
We obtain \cite{ramkumar2022}, 

	\begin{equation}
	\begin{split}
		\bra{n} \vec{\sigma} \cdot (\hat{k}_{\gamma 1} \times \vec{\epsilon}_1^*)\ a^\dagger\ e^{-i \vec{k}_{\gamma 1} \cdot \vec{r} + i\omega_1 t} \ket{i} &= \\
		e^{i\omega_1 t} (-i\cos\phi_{\gamma 1} - \sin\phi_{\gamma 1}) I_1 
	\end{split}
\end{equation}
where $\phi_{\gamma 1}$ is the azimuthal angle of $\vec k_{\gamma 1}$, 
\begin{equation}
	I_1=\int dr\ U_n^*\frac{\sin ( k_{\gamma 1}r)}{ k_{\gamma 1} r}U_i 
	\label{eq:I1}
\end{equation}
and $U_i$ and $U_n$ are the initial and intermediate state wave functions
respectively. 

We next consider the transition from the intermediate to the final state with the emission of another photon.  
The final nuclear state has $l=1$ with $j=3/2$ and $j_z=3/2$. Hence the 
dominant contribution comes from 
the first term of Eq. \ref{eq:IH}. Let the wave vector of 
the emitted photon be $\vec k_{\gamma 2}$ and frequency $\omega_2$. We denote the polar coordinates of the unit vector $\hat k_{\gamma 2}$ by $\theta_2$ and $\phi_2$. The polarization vectors are denoted by $\vec \epsilon{\,'}_1$ and $\vec \epsilon{\,'}_2$. 
These three unit vectors are given by Eqs.
\ref{eq:kgamma} and \ref{eq:eps1_2} with suitable replacement of angle 
parameters. The polarization vector of the photon produced at this vertex is
taken to be $\vec\epsilon{\,'}_2$.  
The resulting matrix element is found to be \cite{ramkumar2022},
\begin{equation}
	\begin{aligned}
		\bra{f}H_I(t)\ket{n} &= -e^{i\omega_2 t}\, i e \sqrt{\frac{1}{12V \omega_2 \hbar}} \\
		&\quad \times (E_f-E_n)(\sin\phi_2 + i\cos\phi_2) I_2.
	\end{aligned}
\end{equation}

where
\begin{equation}\label{eq:I21}
	I_2=\int dr'\ U_f^*\ r'\ U_n
\end{equation}
We point out that photons with wave vectors $\vec k_{\gamma 1}$ and 
$\vec k_{\gamma 2}$ can also be emitted
from vertex 2 and 1 respectively. Hence, both the molecular and nuclear 
matrix element has an additional contribution. However, for our choice of
parameters, with $k_{\gamma 2}>>k_{\gamma 1}$, the second term can be safely dropped.

The reaction rate is given by, 
\begin{equation}
	{dP\over dt} = {1\over \Delta T} \int dE_1 dE_2 \rho_1\rho_2 
	|\langle f|T(t_0,t)|i\rangle|^2
\end{equation}
where,
\begin{eqnarray}\label{eq:tm1}
	\langle f| T(t_0,t)| i \rangle &=\bigg(\frac{-i}{\hbar}\bigg)^2 \sum_{n} \int_{t_0}^{t} dt' \langle f |e^{\frac{i H_0 t'}{\hbar}} H_I(t') e^{-\frac{iH_0 t'}{\hbar}}\nonumber \\
	&|n\rangle   \int_{t_0}^{t'} dt'' \langle n |e^{\frac{i H_0 t''}{\hbar}} H_I(t'') e^{-\frac{i H_0 t''}{\hbar}}|i\rangle
\end{eqnarray}	
is the transition amplitude of the process at second order 
\cite{merzbacher1998quantum,sakurai1967advanced}. 
Here $E_1=\hbar \omega_1$, $E_2=\hbar \omega_2$ and 
$\rho_1$ and $\rho_2$ are the photon density of the state factors.
We obtain \cite{ramkumar2022}
\begin{equation}
	\frac{dP}{dt} = \frac{\alpha^2 g_p^2}{12\pi\hbar^3 c^6m_p^2}\int dE_1\ E_1^3\ E_2\ |I|^2
	\label{eq:dPdt}
\end{equation}
where,
\begin{equation}\label{eq:inter}
	I=\sum_nI_1I_2\frac{E_f-E_n}{E_n-E_i+E_1}
\end{equation}
In Eq. \ref{eq:inter}, the sum is over all the intermediate states 
and $E_2 = E_i-E_f-E_1$. 
The energy eigenvalues are continuous and we convert the sum into an integral, 
\begin{equation}\label{eq:sumint}
	\sum \rightarrow L \int \frac{dk_n}{\pi}
\end{equation}
where $L\rightarrow \infty$ and cancels out of the calculation. 
Hence, we obtain
\begin{equation}\label{eq:inter1}
	I={L\over \hbar \pi}\sqrt{m_p\over 2}\int_0^\infty {dE_n\over \sqrt{E_n}}\,  I_{1} I_{2} \frac{E_{f}-E_{n}}{E_{n}-E_{i}+E_{1}}
\end{equation}
The cross section can be evaluated as in the case of the first order
process. For the initial state we use the plane wave normalization
and set the corresponding normalization factor to be $N_i = k_iL^{3/2}\sqrt{D_i^2
	+F_i^2}$, where $D_i$ and $F_i$ are the wave function coefficients 
corresponding to the initial state.

\subsection{Calculations}
\label{sec:calculations}

We are interested in contribution from a resonant state with energy $E_n<E_i$. 
The dominant contribution is obtained for $E_n\approx E_i-E_1$. At this precise
value, the denominator in Eq. \ref{eq:inter} becomes zero and the integral 
is ill defined. We assign a small width to the intermediate state and replace
$E_n$ by $E_n-i\Gamma/2$, with $\Gamma$ taken to zero at the end of the 
calculation. Hence, we may write
\begin{equation}
	{1\over E_{n}-E_{i}+E_{1}-i\Gamma/2}= {E_{n}-E_{i}+E_{1}+i\Gamma/2 
		\over (E_{n}-E_{i}+E_{1})^2 + \Gamma^2/4}
\end{equation}
In the limit $\Gamma\rightarrow 0$ the imaginary part can be approximated
as a delta function. Hence, this part of the integral over $E_n$ can be
easily evaluated and the real part adds incoherently to the final cross section.
Here we focus on this integral since it will provide a lower limit 
on the cross section. We set 
\begin{equation}
	\lim_{\Gamma\rightarrow 0}	{\Gamma/2\over (E_{n}-E_{i}+E_{1})^2 + \Gamma^2/4} = \pi\delta(E_{n}-E_{i}+E_{1})
	\label{eq:deltafn}
\end{equation}
We perform the integral in Eq. \ref{eq:inter1} by extending the lower limit
to $-\infty$. 
This leads to 
\begin{equation}\label{eq:inter11}
	I={iL}\sqrt{m_p\over 2}{1\over \hbar\sqrt{E_n}}\,  I_{1} I_{2} 
	(E_{f}-E_{n})\Bigg|_{E_n=E_i-E_1}
\end{equation}
where $E_n=E_i-E_1$.
We next need to evaluate the integral over the photon energy given in 
Eq. \ref{eq:dPdt}. 

Let us denote the wave function coefficients, Eq. \ref{eq:coeff},
for the state $U_n$ by
$B_n$, $C_n$, $D_n$ and $F_n$. 
The dominant contribution to cross section would come from a
very small region for which $C_n\approx 0$. Let us denote the energy 
corresponding to this by $E_n=E_R$. We expand all quantities around $E_n=E_R$ 
and drop terms high order in $E_n-E_R$. 
We obtain
\begin{equation}
	C_n = \left(E_n-E_R\right) \tilde C + ...
\end{equation}
where 
\begin{equation}
	\tilde C = {e^{-\kappa_{2R}L_n} \sqrt{m_p}\over 2\sqrt{2}\hbar \kappa_{2R}}
	\left({C_1\over \sqrt{V_0+E_R}} - {\sin k_{1R}L_n\over \sqrt{V_1-E_R}}
	\right)
\end{equation}
and
\begin{equation}
	C_1 = \kappa_{2R} L_n\cos k_{1R}L_n
	-k_{1R}L_n\sin k_{1R}L_n + \cos k_{1R}L_n
\end{equation}
Here $\kappa_{2R}$, $k_{1R}$ and $B_R$ refer to the values of $\kappa_2$,
$k_1$ and $B_n$ respectively at the intermediate state energy $E_n=E_R$.
The coefficient $B_n$ can directly be computed at $E= E_R$ and
\begin{eqnarray}
	D_n &=&  (E_n-E_R)\tilde C e^{\kappa_{2R}L_b} \left[\sin k_R L_b
	+{\kappa_{2R}\over k_R}\cos k_RL_b\right] \nonumber\\
	&+& B_Re^{-\kappa_{2R}L_b} \left[\sin k_R L_b
	-{\kappa_{2R}\over k_R}\cos k_RL_b\right]
	\nonumber\\
	F_n &=&  (E_n-E_R)\tilde C e^{\kappa_{2R}L_b} \left[\cos k_R L_b
	-{\kappa_{2R}\over k_R}\sin k_RL_b\right]\nonumber\\
	&+& B_Re^{-\kappa_{2R}L_b} \left[\cos k_R L_b
	+{\kappa_{2R}\over k_R}\sin k_RL_b\right]
\end{eqnarray}
Furthermore, we find,
\begin{equation}
	\begin{split}
	{1\over D_n^2+F_n^2} &= {e^{-2\kappa_{2R L_b}}/\tilde C^2\over
		\left(E_n-E_R+\Gamma_1\right)^2 + \left(\kappa_{2R}\over k_R\right)^2
		\left(E_n-E_R-\Gamma_1\right)^2}
	\end{split}
	\label{eq:deno}
\end{equation}
where 
\begin{equation}
	\Gamma_1 = {B_R\over \tilde C} e^{-2\kappa_{2R L_b}}
\end{equation}

The integral $I_1$ can be expressed as,
\begin{equation}
	I_1 = I_{11}+I_{12}+I_{13}+I_{14}+I_{15}+I_{16}
\end{equation}
where
\begin{eqnarray}
	I_{11} &=& {B_nC_i\over N_nN_i}\int_{L_n}^{L_b} dr e^{-(\kappa_{2R})r}
	{\sin k_\gamma r\over k_\gamma r} e^{\kappa_{2i}r}\nonumber \\
	I_{12} &=& {C_nC_i\over N_nN_i}\int_{L_n}^{L_b} dr e^{(\kappa_{2R})r}
	{\sin k_\gamma r\over k_\gamma r} e^{\kappa_{2i}r}\nonumber \\
	I_{13} &=& {D_nD_i\over N_nN_i}\int_{L_b}^{\infty} dr \sin{k_Rr}
	{\sin k_\gamma r\over k_\gamma r} \sin{k_ir}\nonumber \\
	I_{14} &=& {D_nF_i\over N_nN_i}\int_{L_b}^{\infty} dr \sin{k_Rr}
	{\sin k_\gamma r\over k_\gamma r} \cos{k_ir}\nonumber \\
	I_{15} &=& {F_nD_i\over N_nN_i}\int_{L_b}^{\infty} dr \cos{k_Rr}
	{\sin k_\gamma r\over k_\gamma r} \sin{k_ir}\nonumber \\
	I_{16} &=& {F_nF_i\over N_nN_i}\int_{L_b}^{\infty} dr \cos{k_Rr}
	{\sin k_\gamma r\over k_\gamma r} \cos{k_ir}
\end{eqnarray}
Here we have set the intermediate state momentum $k_n$ equal to its value
at resonant energy $k_R$. The integral $I_2$ can be expressed as,
\begin{eqnarray}
	I_2 &=& {1\over N_n} \int_0^{L_n} dr U_f r \sin k_Rr\nonumber \\
	&+& {B_n\over N_n} \int_{L_n}^{L_b}dr U_f r e^{-\kappa_{2R}\,r}\nonumber\\
	&+& {C_n\over N_n} \int_{L_n}^{L_b} dr U_f r e^{\kappa_{2R}\,r}
\end{eqnarray}

The integral over the photon energy $E_1$ given in 
Eq. \ref{eq:dPdt} can now be evaluated by setting the intermediate
energy $E_n=E_i-E_1$ in all integrals over $r$ at its resonant value $E_R$.
Hence, we set a fixed value of $E_1$ in 
all the terms which vary slowly with $E_1$.  
The integral over photon energy $E_1$ then involves integrals such as
$$\int dE_1 {D_n^2\over \left(D_n^2+F_n^2\right)^2} \ \ \ \ {\rm and}
\ \ \int dE_1 {F_n^2\over \left(D_n^2+F_n^2\right)^2} \,.
$$
Here the factor $1/(D_n^2+F_n^2)$ arises from the factor $1/N_n$ in the 
integral $I_2$ and the remaining factor arises from the integrals which
contribute to $I_1$.
The integrand gets dominant contribution from a very small interval 
$\Delta E_1$ of order $e^{-2\kappa_R L_b}$ 
of the photon energy. Within this interval the integrand is of order 
$e^{2\kappa_R L_b}$. Hence, we expect a finite and significant
result for the integral.

The factor $1/(D_n^2+F_n^2)$, given in Eq. 
\ref{eq:deno}, gets dominant contribution from $E_n-E_R\approx \Gamma_1$. 
This is because $\kappa_{2R}>> k_R$. We express this factor as,

\begin{equation}
		\begin{aligned}
	{1\over D_n^2+F_n^2} &\approx {k_R^2e^{-2\kappa_{2R L_b}}\over \kappa^2_{2R}\tilde C^2} \times\\
	&{1\over	\left(E_n-E_R-\Gamma_1\right)^2 + \left(k_{R}\over \kappa_{2R}\right)^2
		\left(E_n-E_R+\Gamma_1\right)^2}
	\end{aligned}
\end{equation}

Since the second term in the denominator contributes only when 
$E_n-E_R\approx \Gamma$ we can set $E_n-E_R=\Gamma_1$ in this term. This factor
behaves as a delta function and we can approximate it as,
\begin{equation}
	{1\over D_n^2+F_n^2} \approx {k_R\over 2\kappa_{2R}|\tilde CB_R|}
	\pi \delta (E-E_R-\Gamma_1)
\end{equation}
where we have explicitly substituted for $\Gamma_1$. Integration over photon
energy $E_1$ can now be carried out using this delta function. It is clear
that now all the large exponential factors have cancelled out. We point out 
that the factors, such as, $D_n^2/(D_n^2+F_n^2)$ are order unity over the
entire range of energy. 
We express
\begin{equation}
	I_2^2 = N_n^2 \tilde I_2^2
\end{equation}
and carry out the integral over $E_1$. This leads to 
\begin{equation}
	{dP\over dt} = {\alpha^2 g_p^2L\over 24\hbar^5 c^6m_p}{(E_f-E_R)^2
		E_1^3E_2 k_R\over E_R\kappa_{2R} |\tilde CB_R|} \tilde I_2^2 I_1^2
\end{equation}
where $E_R$ is the energy of the resonance and $E_1 = E_i-E_R$.   
Finally, the cross section is given by,
\begin{equation}
	\sigma = {L^3\over v} {dP\over dt} 
\end{equation}
where $v=\sqrt{2E_i/m_p}$ is the initial state velocity. 
The term $I_1^2$ contains a factor $1/L^4$ arising from the normalization, 
which will cancel with the $L^4$ factors in the numerator.

\subsection{Results}
\label{sec:results}
We use the following potential parameters:
\begin{eqnarray}
	V_0 &=& 5.0\times 10^7\ {\rm MeV}\nonumber \\
	V_1 &=& 100 \ {\rm atomic \ units}\nonumber\\
	L_b &=& 0.1 \ {\rm atomic \ units}\nonumber \\
	L_n &=& 0.574585\times 10^{-4}  \ {\rm atomic \ units} 
\end{eqnarray}
The value of $L_n$ is adjusted so that we may obtain a resonance at low
energy. With this choice of $L_n$ we obtain resonance energy $E_R=0.53869$
atomic units. We obtain the nuclear state corresponding to $l=1$
first radial excitation at $E_f = -13.56$ MeV. The initial state is taken
to be at energy 1 atomic unit, i.e. 27.2 eV. With these parameters, we find
the cross section to be equal to $7\times 10^{-12}$ barns, which is sufficiently large
to be experimentally observable. 
In comparison, the first order transition with initial energy 1 atomic unit
leads to the cross section equal to $9.5\times 10^{-41}$ barns.  
The main point is that, in contrast to the
second order resonant process,  
this is heavily suppressed by the exponential factor $\exp(-2\kappa_2L_b)$
which decays sharply with increase in the barrier potential. However, 
increase in barrier potential
has a relatively mild effect on the second order process.

In our analysis, we have assumed that the target nucleus X has very large 
mass. Including contributions from mass of this nucleus simply introduces
a small change in the over all factor and changes the energies of the produced
photons by a small amount due to the energy carried by the nuclear recoil. 
The generalization to Coulomb and more refined nuclear 
potentials is mathematically
a little more complicated 
but we do not expect any essential change in our results. 
The first important step is the integral over intermediate state energy
$E_n$ which again proceeds by the delta function approximation given in 
Eq. \ref{eq:deltafn}. We next need to perform the integral over the 
photon energy $E_1$ in Eq. \ref{eq:dPdt}. In this case the integrand 
involves two matrix elements $I_1$ and $I_2$. Here $I_1$ represents the
transition from 
the initial state to intermediate state and $I_2$ the transition from 
intermediate to final state. Both of these integrals are expected to be
non-zero and their strengths depend on the actual nature of the transition. 
We again expect a very large contribution from a very small region of 
$E_1$ corresponding to the width of the state and negligible beyond that. 
The width is directly related to the exponential suppression factor
and hence we expect these factors to cancel out in the integral over 
$E_1$ leading to a final result which does not depend on such factors.
We postpone this 
as well as applications to realistic nuclei to future research.

\section{Conclusions}
We have studied the process of fusion of a proton with a nucleus X at 
very low initial energy using a toy step model for the repulsive potential.
We have shown that the cross section for this process is significantly
large and observable in the presence of a low energy resonance. This is
applicable for a second order process which goes through two interactions.
In our analysis we have assumed both of these to be electromagnetic and 
the process leads to production of two photons. 
We also find that a related first order process, which leads to emission
of only one photon, has very large cross section for a very narrow range
of energies. Due to the very narrow energy range, this process does not
lead to a significant rate in any experimental situation. In contrast, the
second order process does not require any fine tuning of the initial state
energy and leads to a significant rate as long as the initial state
particles have energies larger than the resonant energy.
	
	\bibliographystyle{ieeetr}
	\bibliography{nuclear}

\end{document}